\documentclass[sn-mathphys,Numbered]{sn-jnl}

\usepackage{graphicx}%
\usepackage{multirow}%
\usepackage{amsmath,amssymb,amsfonts}%
\usepackage{amsthm}%
\usepackage{mathrsfs}%
\usepackage[title]{appendix}%
\usepackage{xcolor}%
\usepackage{textcomp}%
\usepackage{manyfoot}%
\usepackage{booktabs}%
\usepackage{algorithm}%
\usepackage{algorithmicx}%
\usepackage{algpseudocode}%
\usepackage{listings}%

\theoremstyle{thmstyleone}%
\theoremstyle{thmstyletwo}%

\theoremstyle{thmstylethree}%

\begin{document}

\title[Service Composition in the ChatGPT Era]{Service Composition in the ChatGPT Era}

\author{\fnm{Marco} \sur{Aiello}}\email{marco.aiello@iaas.uni-stuttgart.de}

\author{\fnm{Ilche} \sur{Georgievski}}\email{ilche.georgievski@iaas.uni-stuttgart.de}

\affil{\orgdiv{Service Computing Department, IAAS},
  \orgname{University of Stuttgart},
  \orgaddress{\street{Universit\"atsstr. 38}, \city{Stuttgart}, \postcode{70569}, \state{BW}, \country{Germany}}}


\maketitle


ChatGPT recently attracted vast attention in and outside the research community for its conversational abilities that mimic human ones exceptionally well. At the heart of systems like ChatGPT are Large Language Models (LLM). These models, rooted in deep neural networks, have the ability to predict the next textual token in a series of tokens based on statistical occurrences in extremely large data sets~\cite{wolfram2023chatgpt}. When the models are sufficiently big and well-tuned, one observes the ``unreasonable effectiveness of data''~\cite{halevy2009unreasonable} in how the system generates perfectly intelligible and believable sentences. Such ability to have human-like conversations with a software system is both stunning for the quality of the conversation and mind-blowing in terms of the potential impact on society and the job market in particular~\cite{eloundou2023gpts,smi:job23}.
There is controversy on whether or not such systems manifest forms of artificial intelligence. Researchers at Microsoft attribute signs of intelligence to the current fourth version of Generative Pre-trained Transformer (GPT-4), which is in development at the time of writing~\cite{bubeck2023sparks}. Some authors have successfully solved Theory of Mind tasks using such tools. Kosinski reports a success rate of 95\% using GPT-4 in solving false-belief tasks. Other authors are more careful with the excessive anthropomorphization of ChatGPT-like systems~\cite{shanahan2023talking,bro:gpt23}.
What is sure is that the embedding of an LLM into a system makes it a very powerful tool.
Of interest to us in this editorial is the LLM capability to generate programming code~\cite{mey:gpt22} and its potential impact on Service-Oriented Computing and Applications.

The problem of automated service composition is central to the field of Service-Oriented Computing and Applications~\cite{papazoglou2003service}. Seamless, unsupervised, automated composition of services available on a network is a potent way to build adaptive information systems. It is the idea that one can execute any task relying on multiple, loosely-coupled services, possibly without prior knowledge of such services. This would guarantee that virtually any task can be executed by relying on third-party implementations and resources. 

Such a vision of automation is rooted in the fields of software engineering and component-based software engineering. It emerged in a fervent moment of technological evolution. At the beginning of the century, the Internet was becoming pervasive, and the Web emerged as a central technology also for businesses. Every company was moving towards having a Web presence, which also meant having a web server with a port always open as a gateway to their enterprise information systems~\cite{aie:web18}. Standardized interoperation was also emerging as the Web language HTML was being stripped of its layout and presentation aspects and turned into a platform-agnostic data description language, XML. Specific dialects of XML were proposed to describe wiring protocols for remote interoperation (SOAP), for describing services (WSDL), for describing service orchestrations (BPEL), and service discovery infrastructure was also proposed (UDDI)~\cite{wsbook2005}. The core pattern behind using such technology is that services are published, then they can be discovered, and once found, they can be invoked {\em (publish-find-bind).}

The technological availability of XML-described services inspired many researchers to propose techniques for achieving automated service composition, see~\cite{dustdarS05} for an early survey of the systems. One of the biggest challenges that emerged from the beginning was the necessity to understand what a service was capable of by simply looking at the signature of its interfaces. While the format and type of the exchange data are declared, the service's actual implementation and business goals are unknown to the service composition engine. This means that either the proposed systems were just doing syntactic matching of the interfaces but could not really provide guarantees of what the composition would do, or some additional semantic information regarding the services had to be (manually) provided. Both the syntactic and the semantic approaches would prove to have drawbacks that made the proposals inapplicable in practice. Though, the road of providing additional semantics to the service description was the more intriguing and practiced one. The very strong assumption is that all service implementations come with additional descriptions of the preconditions for running a service and the outcome of invoking the service. In other words, if one invokes a weather service, a time and place need to be provided, and in the end, one will know the temperature in that place at that time. Notice that this is much more information than simply asking to provide three inputs, two floats and a string, and receive back as output an integer.  
\begin{figure}[htb]
	\centering
	\includegraphics[width=\textwidth]{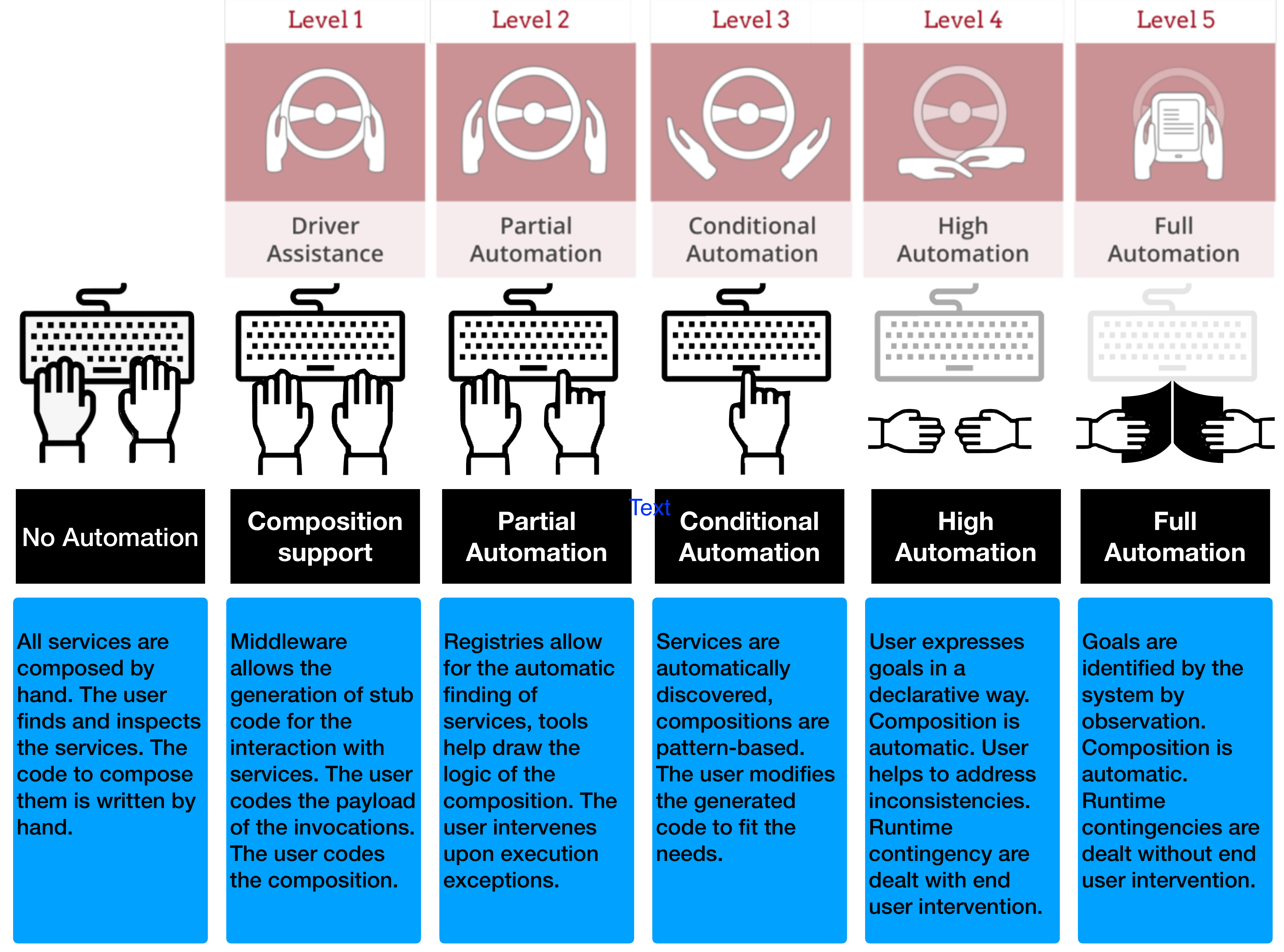}
	\caption{Maturity levels of Automated Service Composition,
          from~\cite{DBLP:conf/icsoc/Aiello22}}
	\label{fig:levels}
\end{figure}
While semantic service annotations describing the inner working of the services would have made automated service composition much more realistic, the truth is that the world went in another direction and such descriptions were never broadly available. It was not the precise semantic descriptions that would come but a huge amount of data that, taken jointly, would be statistically meaningful and would provide enough information for performing basic tasks. This is how the first versions of the Google search engine become successful, not by understanding the content of the webpages, but by looking at the frequency of words and link structure of the Web. The promises of ChatGPT today come from the same place of data abundance and statistical relevance.\footnote{Incidentally, the fortune of the Semantic Web followed a similar destiny to that of automated service composition for basically the same reasons. The Semantic Web is the technological evolution of the Web where textual content is augmented with semantic annotations~\cite{aie:web18}. The annotations refer to formally defined ontological information on which one can reason. The idea behind its definition was to add descriptions of the meaning of terms to the Web so that more precise searches could be performed and potentially also reasoning. But again, the world did not find enough value in adding such descriptions and the Semantic Web technology was never widely adopted.}

\begin{figure}[htb]
	\centering
	\includegraphics[width=130mm]{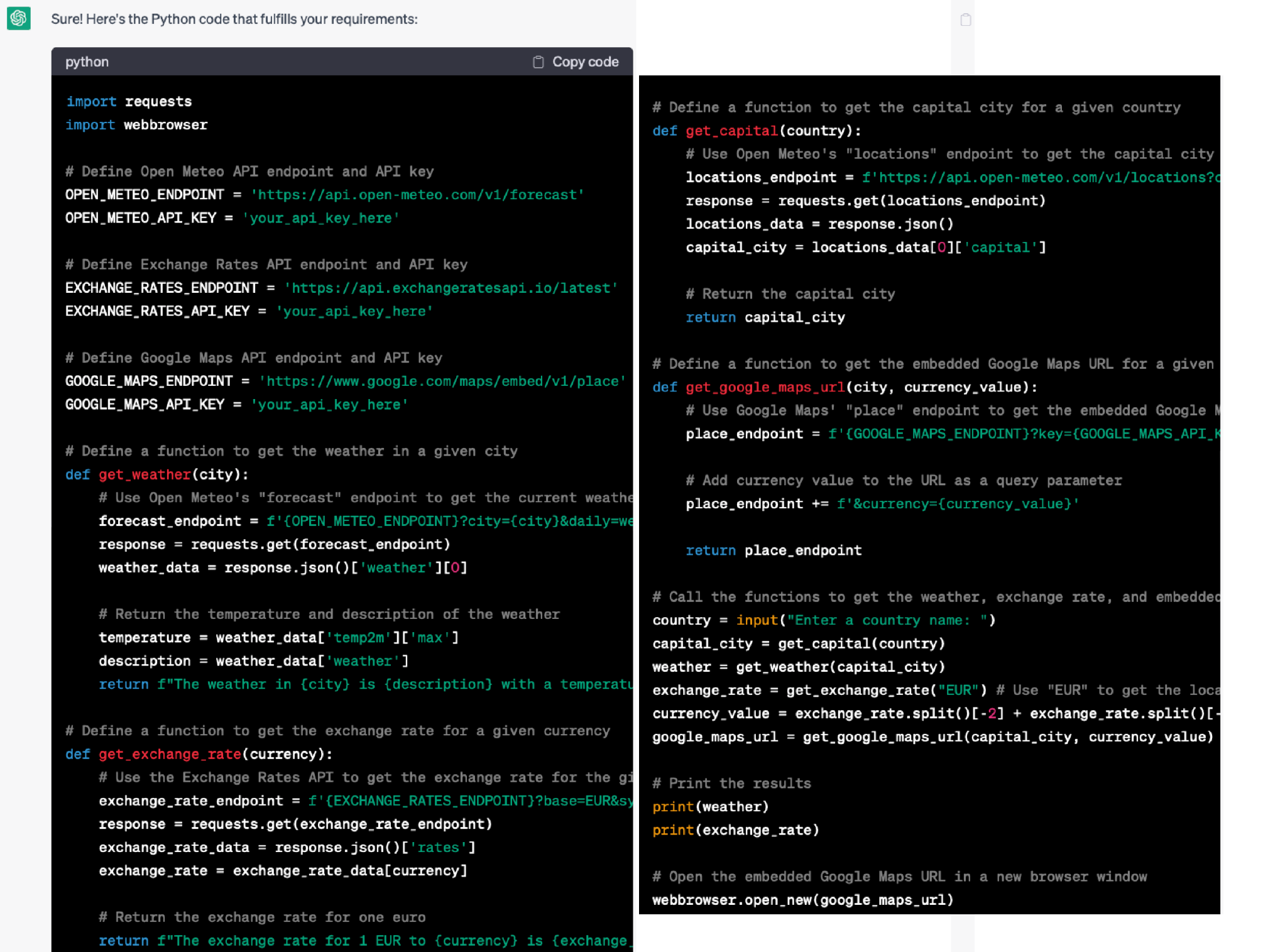}
	\caption{Generated Python service composition code, by GPT-3.5}
	\label{fig:code}
\end{figure}

There is no point to go against technological evolution. The approach of adding semantic information to service descriptions (and web pages via the Semantic Web, for that matter) was never successful and was superseded by the simple abundance of data, directly or indirectly generated by humans. The question then becomes: can such data abundance help in automated service composition?

In the vision paper about service composition for the next 50 years,
we draw a parallel between the levels of autonomous driving and the
development of autonomous service
composition~\cite{DBLP:conf/icsoc/Aiello22}. The reasoning is that,
similarly to autonomous driving, the automation of service composition
goes through levels and needs technological innovations to 
advance from one step to another. Machine learning and computer vision
advancements have helped cars achieve ``Partial Automation,'' i.e.,
Level 2. For automated service composition, we have been long stuck
with systems of Levels 1-2, and after the initial results of a couple
of decades ago, not much innovation has emerged. LLM promises to be a
technological innovation that can revive the field and bring it from
Level 2 to higher levels of automation. Let us look at an example to
see what is already possible today with an available tool such as ChatGPT.

\begin{figure}[htb]
	\centering
	\includegraphics[width=100mm]{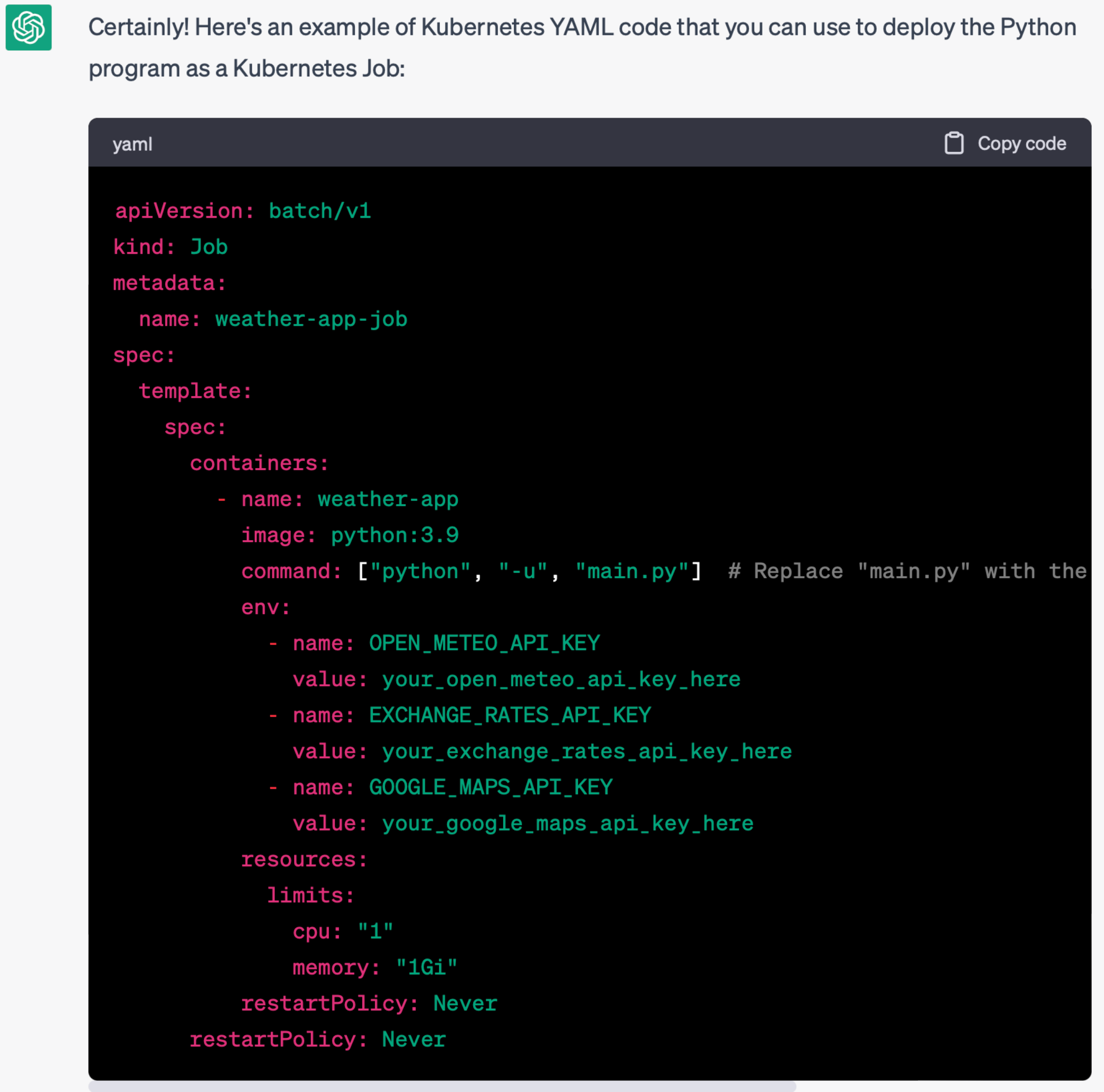}
	\caption{Generated Kubernetes script, by GPT-3.5}
	\label{fig:deploy}
\end{figure}

The classic examples of service composition have to do with traveling around the world. 
Say that one wants a composition that given a country provides the weather in its capital city and the currency conversion rate to the Euro. This requires interaction with several Web services about weather, geography, and currency conversion. It requires passing parameters around that have to fit the syntax and data types of each service and requires knowing that a country has a capital and a currency. The traditional approach to solve this composition is for an engineer to search for the available services, explore their APIs, and then write some orchestration of the services in a workflow language or in some programming language to be run in an application container. This would cost a developer between a few hours and a day, based on his experience and abilities. Could ChatGPT do it for us? 
We could phrase the composition as: ``Given the name of a country, find what the weather is in the capital using the API of https://open-meteo.com and what is the value of the local currency for one euro, and show the city with the currency value on Google Maps via the Google Maps API. In Python, please.''

If we use GPT version 3.5 (GPT-3.5) and provide it the text in quotes above, we do get Python code! This is both surprising and impressive. The code, generated on May 10th, 2023,  is displayed in Figure~\ref{fig:code} and it is generated in less than a minute. It looks ready for execution, though it does need editing, most notably, one has to add the API keys for the identified services and create a container for running the code.  
In other terms, GPT-3.5 has automated an important number of aspects related to composition. It has performed service discovery. The query did mention where to find the map and weather services but not where to find the currency conversion service. The system has interpreted the interfaces and was able to create input and output parameters. It did even more; it created a function to extract the text identifying the capital of a country given the country's name. Finally, it has created the `business logic' by ordering the invocations and returning both the values and an invocation of the web browser with the requested results. There is though a {\em small} problem: it has invented some parts of the APIs.
For Open-Meteo, no API key is required, while the generated code requires and appends an API key to the endpoint URL. Furthermore, the weather API requires a location to be specified using longitude and latitude coordinates, while the generated code appends a city name to the endpoint URL. The rest of the endpoint URL is also faulty. In summary, ChatGPT generated an endpoint URL with non-existing, faulty, and incomplete parameters. Looking at the generated code responsible for handling the message returned by the weather API, it correctly assumes that the message is in a JSON format, however, the code tries to parse wrong or non-existing fields in the received JSON message. Notice that ChatGPT generated code for acquiring a capital city of the given country. It does so by a call to an Open-Meteo's endpoint that does not exist (see the {\tt get\_capital(country)} function). The code even tries to extract the name of the capital from a JSON message. Similar observations can be made for the other APIs.

While it is impressive that ChatGPT was able to discover services to perform the composition, in our case, Exchange Rates API, that it could ``understand'' the logic of the request and generate the Python code for the composition, that it ``understood'' how to send requests to the services; it has difficulties interpreting the interfaces of the services by using wrong and non-existing parameters, and has made a call to a non-existing API,
a phenomenon often called {\em hallucination.} We can engage in a chat with the system to try to refine the generated code to be correct. The process is somewhat frustrating as it requires many iterations to only fix the problem partially. For example, we needed to confabulate with ChatGPT just to let it know that Open-Meteo cannot work with a city name and needs a city's longitude and latitude coordinates instead. Eventually, ChatGPT corrected this by engaging  a fourth service, namely OpenStreetMap, to get the coordinates of a given city. We can also simply correct the code manually and have the composition ready for execution. The next step is to deploy the code and let it orchestrate the services. We can resort to ChatGPT also for this task by stating something like ``Can you write Kubernetes code to deploy the above Python program?'' and we get the mostly correct Kubernetes YAML code presented in Figure~\ref{fig:deploy} together with instructions on how to customize the YAML code and the Unix shell commands to run (for brevity reasons not shown here). 

This short experience shows that ChatGPT is capable of service discovery, generating compositions, and service interface interpretation. This is a very powerful automation that potentially eliminates the need for service registries (e.g. UDDI), composition languages (e.g., BPEL), and interface description languages (e.g. WSDL). It also shows that it can take care of the deployment aspects of the code, potentially eliminating the need for workflow engines. It also shows that the output is not entirely correct and iterations and supervision by a domain expert are still necessary. Going back to the maturity level diagram shown in Figure~\ref{fig:levels}, ChatGPT shows very high potential to boost the state of the art beyond Level 2 and be at the heart of tools that can automate the creation of service-oriented systems with some small supervision. It shows that ChatGPT can be the type of tool that will move automated service composition from its current state to Automation Level 3.  

To meet such promise, a number of research directions emerge. These are new to the field of automated service composition.
\begin{itemize}
\item{\bf Prompt engineering.} What is the best way to formulate a service composition request? If it is natural language, how should this be done effectively? And how to refine such requests through interaction in order to possibly converge to a correct output? Can one integrate natural language, formal specifications, and programming languages into one request? Is there a way the system can ask back questions to the user in order to resolve ambiguities or composition alternatives?
\item{\bf Composition verification and testing.} Compositions generated using a formal language approach are correct by design (e.g.~\cite{kal:pla16}). This is not true for compositions generated by ChatGPT. How does one perform verification and testing of compositions generated by LLM-like tools? How can one ensure that the composition is satisfying the original request of the user, especially if expressed in natural language? 
\item{\bf Execution monitoring.} Once the composition is deployed and in execution, how can one monitor that it is actually performing as intended? How can it deal with run-time contingencies and non-deterministic and Byzantine service behaviors? These are general problems of composition executions, the new aspect here is to investigate if LLM can be also a tool to address these issues, and if so, how.
\end{itemize}

After the enthusiasm at the beginning of the century about automating service composition, the field has slowed down, and not much progress has been made in the last decade. We believe that the emergence of tools like ChatGPT will have an important impact on automated service composition and its actual practical use, if the open research challenges, including the ones listed above, are addressed. More research and development will show whether the claim of moving soon to Level 3 is correct or not. As a final note, we state that, except for the output screenshots shown in Figures~\ref{fig:code} and~\ref{fig:deploy}, no text of this article has been generated by ChatGPT, but it is entirely the responsibility--for good or for worse--of the authors.

\bibliography{aiello,chatgpt}

\end{document}